# Impact of the Net Neutrality Repeal on Communication Networks

School of Electronic and Electrical Engineering, University of Leeds, LS2 9JT, United Kingdom
Hatem A. Alharbi, Taisir E.H. Elgorashi and Jaafar M.H. Elmirghani

*Abstract*— Network neutrality (net neutrality) is the principle of treating equally all Internet traffic regardless of its source, destination, content, application or other related distinguishing metrics. Under net neutrality, ISPs are compelled to charge all content providers (CPs) the same per Gbps rate despite the growing profit achieved by CPs. In this paper, we study the impact of the repeal of net neutrality on communication networks by developing a techno-economic Mixed Integer Linear Programming (MILP) model to maximize the potential profit ISPs can achieve by offering their services to CPs. We focus on video delivery as video traffic accounts for 78% of the cloud traffic. We consider an ISP that offers CPs different classes of service representing typical video content qualities including standard definition (SD), high definition (HD) and ultra-high definition (UHD) video. The MILP model maximizes the ISP profit by optimizing the prices of the different classes according to the users demand sensitivity to the change in price, referred to as Price Elasticity of Demand (PED). We analyze how PED impacts the profit in different CP delivery scenarios in cloud-fog architectures. The results show that the repeal of net neutrality can potentially increase ISPs profit by a factor of 8 with a pricing scheme that discriminates against data intensive content. Also, the repeal of net neutrality positively impacts the network energy efficiency by reducing the core network power consumption by 55% as a result of suppressing data intensive content compared to the net neutrality scenario.

*Index Terms*— Net neutrality, AT&T, IP over WDM networks, profit, power consumption.

## I. INTRODUCTION

Network (net) neutrality regulations prohibit ISPs from applying different treatment to IP packets based on their content e.g. prioritizing, blocking or throttling certain Internet content or allowing quality differentiation. Net neutrality, which was scrapped by the US Federal Communications Commission (FCC) in December 2017, has been the subject of remarkable debate in recent years between ISPs and CPs with each side trying to exploit their assets and expand their profit and influence. The debate is fueled by the rapidly escalating demand for CPs services as a result of the interconnection between Internet and broadcasting markets. Cisco forecasts [1] that by 2021, annual global Internet traffic will hit 2.2 Zettabytes per month and CPs datacenters will be the source of 71% of this traffic. Online video services are the primary cause of this accelerated growth in Internet traffic. Video streaming is poised to consume 78% of the total CPs bandwidth with 75% of Internet video traffic originating from higher video services quality (HD and UHD).

Proponents of preferential treatment of Internet traffic complain that the increasing demand for data-intensive content creates a significant burden on the communication network. They argue that removing net neutrality will give ISPs further control of their infrastructure, which is crucial in order to improve QoS and reduce security threats. Another argument is that a significant fraction of the profit of this tremendously growing market is seized by CPs whereas ISPs act as a transit or transport medium into CPs customers. In the US, the quarterly profit margin of AT&T (an ISP) has been almost stable over the last six years whereas Netflix (a CP) profit margin has risen up in rapid pace from 0.7% to 9.8% within the same period [2], [3]. In contrast, advocates warn that removing net neutrality will slow down the innovation in the Internet and its content and will limit the content competition by disadvantaging small businesses, and subsequently, diminish online services.

Deploying traffic discrimination in video delivery services has many challenges, e.g. detecting video packets and enforcing a policy on a certain video quality. Traffic discrimination in IP communication networks has been surveyed intensively in the literature. Several traffic management practices have been surveyed in [4]. The authors highlighted that traffic discrimination taxonomy has four features: (i) characteristics or condition of the traffic (e.g. based on content, protocol or source/destination); (ii) traffic classification (e.g. based on flow rate, header information or routing); (iii) mechanism of discrimination (e.g. modify, delay, drop or block); and (iv) perceived discrimination by end-users. Video traffic can be analyzed using two mechanisms; deep packet inspection (DPI) [5], [6] or traffic profiling [7], [8]. DPI examines the data packets that are sent over the network and traffic profiling detects abnormal network traffic by comparing new traffic against previous traffic profile. For example, an alarm can be triggered if the data rate transmitted over the network (measured in bps) spikes above the desired data rate, which could indicate an increase in data rate. QoS for video services delivery can be applied either by reserving network bandwidth for video packets (e.g. using IntServ) or labelling video content as high priority e.g. by applying Differentiated Services (e.g. using DiffServ) [9].

The Internet ecosystem is complex with many stakeholders. As illustrated in Fig. 1, the main stakeholders in the Internet ecosystem are; ISPs, CPs, content delivery networks (CDNs) and end-users. Users pay ISPs a subscription fee to get Internet access and subscribe to CPs (if required) to access their content. CPs subscribe to a CDN to access storage and processing capacity and to deliver their content to customers. CDNs are responsible for sending CPs content at large scale over ISPs network infrastructure, e.g. the CP Netflix collaborates with the CDN Amazon Web Services (AWS) to reach their customers [10]. ISPs play as the key intermediary in the delivery process as they provide the required connectivity between users and content. Most ISPs such as AT&T [11] and Comcast [12] are now providing CDN services in additional to networking

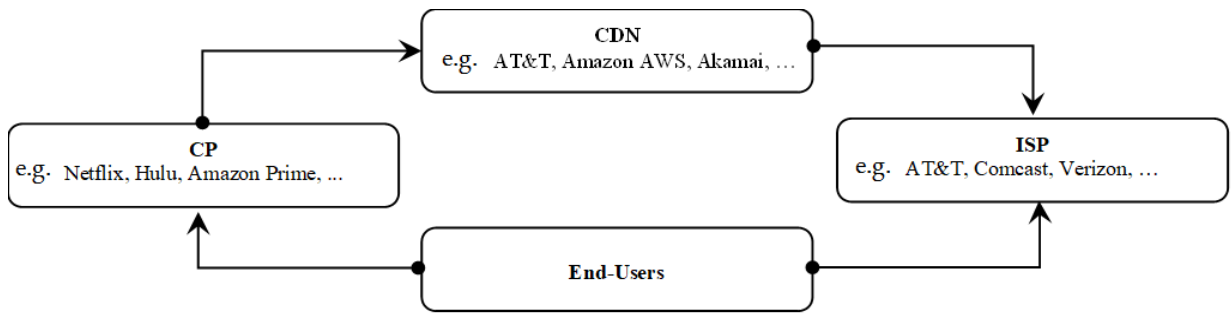

Figure 1: Main stakeholders in Internet ecosystem. Arrows represent customer-provider relationship.

services. To simplify our analysis, we consider a direct relationship between ISP and CPs.

Due to net neutrality regulations, current pricing policy of ISP networking services applies a fixed charge which is not linked with bitrate usage. For example, in the US, AT&T uses a fixed pricing model by charging CPs $3,282 per 10 Gbps per month [13] regardless of the content type transferred to users (either UHD video content or a simple text message).

The rest of this paper is organized as follows: Section II briefly summarizes related work. We describe the pricing scheme we used in this paper and the profit-driven model we adopted in Section III. Our results are presented in Section IV. In Section V, we provide concluding remarks.

## II. RELATED WORKS

Many papers in the literature discussed and analyzed various aspects of net neutrality. From a legalization and regulation perspective, net neutrality in the Internet ecosystem has been surveyed by the authors in [14] and [15]. They emphasized that cloud computing has initiated the net neutrality battle between ISPs and CPs. In [16] the authors analyzed the Internet video streaming contest, taking into account all of ISPs and CPs assets (e.g. content rights, network access, users, …etc). They stated that video distribution makes the dilemmas of net neutrality solid and perceptible. Their analysis demonstrates that net neutrality correlates highly with video service delivery at different points including competition between CPs and ISPs, competition between stand-alone CP and CP owned by ISPs in providing video delivery services and growth of video traffic.

A number of papers in the literature focus on providing mathematical models to investigate the influence of the repeal of net neutrality on communication networks. Paid service differentiation where CPs voluntarily pay a monopoly ISP for prioritizing their traffic under shared network infrastructure was investigated by the authors in [17]. The differentiation occurs where ISPs offer service classes for CPs to choose from where traffic of a higher-priority class will be processed before those of a lower-priority. They studied the optimal pricing based on either maximizing the CPs' choices of service classes or minimizing system delays. Consequently, they highlighted that ISPs optimal pricing strategy can result in an efficient differentiation among CPs maximizing social welfare. Also, they found that applying paid prioritization can lead to money flows (profit) from CPs to ISPs. The authors in [9] modelled the competition of video services delivery market between an ISP's own integrated CP and stand-alone CP. They studied the impact of applying different QoS (marking video traffic as high priority) pricing strategies either by selling QoS to CPs, selling QoS to users, or choosing to not provide QoS at all. They investigated the impact of QoS pricing on the video service prices and CPs profit. The analysis showed that ISPs can sell QoS to CPs at a higher price than when QoS is sold to users, and the CPs are able to make more profit when QoS is directly sold to users than the case when QoS is sold to CPs. Also, they found that an ISP is more likely to use QoS exclusively for its own video services when it provides a similar content of CPs. The cloud infrastructure needed to host and deliver the video content was optimized in [18]-[21] and the impact of the delivery of large data volumes on the network was evaluated in [22]-[25]. Particular attention was paid to the core network which forms the heart of the ISP infrastructure and hosts the CDN with attention given to the network energy efficiency, latency and other QoS metrics [26]-[33]. The work in [34] considered the impact of maximizing profit of CDN providers considering users who access CPs content from either cloud or fog server. In the case of competitive CPs, the CDN always places the content of the popular CP in fog servers, even when a less popular CP pays more, as the CDN tries to reduce core network transit cost.

In this paper, a techno-economic Mixed Integer Linear Programming (MILP) model is built to study the potential profit ISPs can achieve by a differentiated pricing scheme under the repeal of net neutrality. We build on our MILP optimization, network, cloud and fog modelling background [35]-[39] and consider ISPs that offer the CP service classes, which represent different data rate requirements. The model optimizes the pricing scheme of differentiated service classes to maximize the ISP profit based on price elasticity of demand (PED). The MILP model finds the resulting equilibrium pricing, core network power consumption and traffic.

## III. REPEALING NET NEUTRALITY

In this paper, we consider the economic concept of PED to study the impact of ISP's price change on the number of users accessing CPs content. In the following subsections, we present the pricing scheme used in this work followed by the developed network and pricing MILP model.

*A. Pricing scheme:*

In economics, the relationship between users demand and price is referred to as price elasticity of demand (PED) [40]. PED measures the percentage change in demand resulting from one percent change in price. To decide pricing strategy of a product, the seller looks at different sensitivities to various factors that may affect their decision to purchase a product. The dominant factor in determining PED is the users' ability and willingness at any given price. Many factors have an effect on users' behavior such as substitution availability, market competition, frequency of purchase, necessity of the product, and how much the product price represents in users income. The PED is calculated as follows:

$$PED = \frac{\% \text{ Change in Demand}}{\% \text{ Change in Price}} \quad (1)$$

In telecommunications, it is not an easy task to estimate an exact value of PED for various Internet applications as the factors that affect the elasticity change from area to another e.g. wealth, popularity of an application, quality of service provided by ISPs/CPs or competition between different CPs. However, PED for broadband subscriptions in Organization for Economic Co-operation and Development (OECD) countries has been analyzed in [41] by studying the relationship between price, income and broadband adoption. Additional factors have been included in [42], which are age and education to study PED for broadband subscriptions in Latin America and the Caribbean countries. They found that 1% decrease in price would lead to 0.43% and 2.2% increase in demand, respectively, over the two selected areas.

*B. Profit-Driven MILP Model:*

We develop a profit-driven MILP model where the objective is to maximize the total profit of an ISP offering core network infrastructure to CPs to deliver content from distributed clouds and/or fog nodes to their users.

We consider a monopolist ISP who owns the network backbone, i.e. CPs have to subscribe to the monopolist ISP to reach their customers. According to the FCC, 40% of total US Internet subscribers only have a single ISP option in their area [43]. The ISP has the power to control the pricing scheme. Under the net neutrality repeal, the ISP can deliver CPs content of different data rate requirements at a varying price per bit rate. We consider three classes to represent different data rate requirements of CPs services:

- Class A for high data rate content (i.e. UHD video service).
- Class B for medium data rate content (i.e. HD video service).
- Class C for low data rate content (i.e. SD video service).

The ISP needs to optimize the price of the three classes to maximize its profit. We consider content with higher data rate, which causes extra burden on the core network, to be priced higher per bit rate than content with a lower data rate. End-users will perceive varied video definitions from CPs based on their CP subscribed class. We assume that CPs will transfer the ISP new prices to their users to maintain their profit margin. Therefore, for the sake of simplicity we consider CPs to offer the same classes to their users. We assume a certain number of users to initially subscribe to each class under net neutrality. As the ISP and consequently the CPs vary the per bit rate charges for the different classes, users can choose to upgrade, downgrade or unsubscribe to the service. The number of users subscribing to each class depends on the PED. We assume that users leaving class A will join class B, users leaving class B will join class C and users leaving class C will unsubscribe to the service.

The ISP needs to optimize the price of the three classes to maximize its profit. We consider content with higher data rate, which causes extra burden on the core network, to be priced higher per bit rate than content with lower data rate. End-users will perceive varied video definitions from CPs based on their CP subscribed class. We assume that CPs will transfer the ISP new prices to their users to maintain their profit margin. Therefore, for the sake of simplicity we consider CPs to offer the same classes to their users. We assume a certain number of users to initially subscribe to each class under net neutrality. As the ISP and consequently the CPs vary the per bit rate charges for the different classes, users can choose to upgrade, downgrade or unsubscribe to the service. The number of users subscribing to each class depends on the price elasticity of demand (PED). We assume that users leaving class A will join class B, users leaving class B will join class C and users leaving class C will unsubscribe to the service.

Before introducing the model, we define the parameters and variables used in the model:

Parameters:

| | |
|---|---|
| $s$ and $d$ | Indices of source and destination nodes of a traffic demand. |
| $m$ and $n$ | Indices of the end nodes of a physical link. |
| $i$ and $j$ | Indices of the end nodes of a virtual link. |
| $N$ | Set of IP over WDM network nodes. |
| $Nm_m$ | Set of neighbouring nodes of node $m$. |
| α | Set of service classes. |
| $W$ | Number of wavelengths per fibre. |
| $B$ | Wavelength data rate. |
| $CN$ | Number of clouds hosted in core network. |
| $u$ | Total number of users in net neutrality scenario (i.e. before net neutrality is repealed). |
| $LB$ | Minimum percentage of users served by CP to be maintained by the pricing scheme. |
| $d_i$ | Download rate of class $i$. |
| $\epsilon$ | The cost in US$ of provisioning a Gbps of IP over WDM network bandwidth per month. |
| $\ni$ | The cost in US$ of provisioning a Gbps of metro and access network bandwidth per month. |
| $PS$ | The net neutrality selling price in US$ of a Gbps of network bandwidth per month. |
| $E_i$ | Price elasticity of demand of class $i$. |
| $N_{d,i}$ | Number of users of class $i$ located in node $d$ under net |

neutrality scenario.

| | |
|---|---|
| $\delta_s$ | $\delta_s = 1$, if a cloud datacentre is hosted in node $s$, otherwise $\delta_s = 0$. |
| $F_d$ | $F_d = 1$, if there is no fog datacentre hosted in node $d$, otherwise $F_d = 0$. |
| $\zeta$ | Set of all possible solutions. |
| $\rho_{s,i}$ | The price of class $i$ under solution $s$ and class $i$. |
| $yn_{s,d,i}$ | The number of users in solution $s$ subscribing to class $i$ in node $d$ as a result of its PED, where $\frac{PS}{\rho_{s,i} - PS} E_i = \sum_{d \in N} \left( \frac{yn_{s,d,i} - N_{d,i}}{N_{d,i}} \right)$ $\forall i \in \alpha, s \in \zeta.$ |

Variables:

| | |
|---|---|
| $C_{i,j}$ | Number of wavelengths in virtual link $(i,j)$. |
| $W_{m,n}$ | Number of wavelengths in physical link $(m,n)$. |
| $APC_s$ | Number of router ports in node $s$ that aggregate the traffic from clouds. |
| $F_{mn}$ | Number of fibres on physical link $(m,n)$. |
| $L_{i,j}^{s,d}$ | Amount of traffic flow between node pair $(s,d)$ traversing virtual link $(i,j)$. |
| $W_{m,n}^{i,j}$ | Number of wavelengths of virtual link $(i,j)$ traversing physical link $(m,n)$. |
| $r_i$ | ISP's revenue achieved by delivering traffic of class $i$ to CP users. |
| $R$ | Total ISP's revenue in US\$ of delivering networking services to CPs content. |
| $C$ | Total ISP cost in US\$ of provisioning core network. |
| $P_i$ | The price in US\$ per Gbps of network bandwidth per month charged to the class $i$. |
| $U_{d,i}$ | Number of users who subscribe to class $i$ located in node $d$. |
| $CD_{i,d}$ | Cloud flow from users in node $d$ subscribed to class $i$. |
| $Z_{s,i}$ | $Z_{s,i} = 1$, if solution $s$ is selected for class $i$, otherwise $Z_{s,i} = 0$. |
| $ys_{s,d,i}$ | The number of users in solution $s$ subscribing to class $i$ in node $d$, $ys_{s,d,i} > 0$ if solution $s$ is selected for class $i$, otherwise $ys_{s,d,i} = 0$. |

Total ISP's cost and revenue of delivering CP contents are calculated as follows:

Cost of provisioning core, metro and access networks infrastructure ($C$):

$$\sum_{s \in N} APC_s \; B \; \mathcal{E} + \sum_{i \in \alpha} \sum_{d \in N} U_{d,i} \; \Im \; d_i \quad (2)$$

Revenue of delivering networking services to CP users ($R$):

$$\sum_{i \in \alpha} r_i \quad (3)$$

The model is defined as follows:

The objective:

*Maximize total profit given as:*

$$\sum_{i \in \alpha} r_i - \left( \sum_{s \in N} APC_s \; B \; \mathcal{E} + \sum_{i \in \alpha} \sum_{d \in N} U_{d,i} \; \Im \; d_i \right) \quad (4)$$

Equation (4) gives the total profit in US dollar.

The total profit is maximized by maximizing the revenue and minimizing the cost of serving users in different classes.

Subject to:

Revenue of each class:

$$r_i = \sum_{d \in N} U_{d,i} \; d_i \; P_i \qquad \forall \; i \in \alpha \quad (5)$$

Constraint (5) calculates the revenue the ISP achieves by delivering a service class by considering the class price and the total traffic in each class. Note that, the total revenue is obtained by multiplying two variables ($U_{d,i}$ and $P_i$) which is a non-linear process. A look up table of solutions under different PED values defined by parameters $\rho_{si}$, $ys_{s,d,i}$, $yn_{s,d,i}$ is used for linearization. Constraints (6)-(10) select the optimum number of users and price for each class and the calculate the resulting revenue.

$$ys_{s,d,i} \begin{cases} = (yn_{s,d,i} \; Z_{s,i}) & \text{if } i = 1 \\ \leq (yn_{s,d,i} + N_{d,1}) Z_{s,i} & \text{if } i = 2 \\ \leq (yn_{s,d,i} + N_{d,2}) Z_{s,i} & \text{if } i = 3 \end{cases}$$

$$\forall \; i \in \alpha, \; s \in \zeta, d \in N \quad (6)$$

$$P_i = \sum_{s \in \zeta} (\rho_{s,i} \; Z_{s,i}) \qquad \forall \; i \in \alpha \quad (7)$$

$$\sum_{s \in \zeta} Z_{s,i} = 1 \qquad \forall \; i \in \alpha \quad (8)$$

$$U_{d,i} = \sum_{s \in \zeta} ys_{s,d,i} \qquad \forall \; d \in N, i \in \alpha \quad (9)$$

$$r_i = \sum_{s \in \zeta} \left( \sum_{d \in N} ys_{s,d,i} \; d_i \; \rho_{s,i} \right) \qquad \forall \; i \in \alpha \quad (10)$$

Constraint (6) calculates the number of users located in node $d$ of class $i$ in solution $s$. The number of users in class A is the number of users subscribing to the class as a result of its PED (from a look up table). In the case of class B, the number of users available to class B includes all users subscribing to the class B as a result of its PED plus any users downgrading their subscription from class A to class B. In the case of class C, the number of users available to class C includes users subscribing to class C as a result of its PED plus any users downgrading their subscription from class B to class C. Constraint (7) gives the price of each class based on the solution selected from the lookup table. Constraint (8) ensures that only one solution is selected. Constraint (9) calculates the number of users of class $i$ in node $d$. Constraint (10) calculates the revenue the ISP achieves by delivering a service class by multiplying the class price by the total traffic in each class.

Constraints on number of users and prices:

$$\sum_{d \in N} \sum_{i \in \alpha} U_{d,i} \geq u \, LB \tag{11}$$

$$P_1 \geq P_2 \geq P_3 \tag{12}$$

$$\frac{\sum_{d \in N} U_{d,i}}{\sum_{d \in N} \sum_{i \in \alpha} U_{d,i}} = \frac{U_{d,i}}{\sum_{i \in \alpha} U_{d,i}} \tag{13}$$

$$\forall \, i \in \alpha, d \in N$$

Constraint (11) defines the minimum user percentage the CP service needs to maintain. Constraint (12) ensures that the price of a lower class does not exceed the price of upper classes, i.e. the price of class C does not exceed the price of class B and the price of class B does not exceed the price of class A. Constraint (13) ensures that the ratio of users in different nodes is identical.

Core network traffic:

$$\begin{array}{c} CD_{d,i} = U_{d,i} \, F_d \, d_i \\ \forall \, d \in N, \quad i \in \alpha \end{array} \tag{14}$$

$$\sum_{s \in N} L_{s,d} = \sum_{i \in \alpha} CD_{d,i} \tag{15}$$

$$\forall \, d \in N$$

Constraint (14) ensures that nodes with a fog built in its proximity are not served by a cloud. Constraint (15) calculates the download traffic from CP cloud to users in different nodes.

User demands can be used to decide on datacenter locations as follows:

$$L \sum_{d \in N} L_{s,d} \geq \delta_s \qquad \forall \, s \in N \tag{16}$$

$$\sum_{d \in N} L_{s,d} \leq L \, \delta_s \qquad \forall \, s \in N \tag{17}$$

Constraints (16) and (17) relate the binary parameter that indicates whether there is a datacentre built in node $s$ or not ($\delta_s$) to the traffic between users in node $d$ and datacentre in node $s$.

Traffic flow conservation constraint in the IP layer:

$$\sum_{j \in N : i \neq j} L_{i,j}^{s,d} - \sum_{j \in N : i \neq j} L_{i,j}^{s,d} = \begin{cases} L_{s,d} & i = s \\ -L_{s,d} & i = d \\ 0 & otherwise \end{cases}$$

$$\forall \, s, d, i \in N : s \neq d \tag{18}$$

Constraint (18) represents the flow conservation for IP layer in the IP over WDM network. It ensures that the total incoming traffic equal the total outgoing traffic in all node; excluding the source and destination nodes.

Virtual link capacity constraint:

$$\sum_{s \in N} \sum_{d \in N : s \neq d} L_{i,j}^{s,d} \leq C_{i,j} \, B$$

$$\forall \, i, j \in N : s \neq d \tag{19}$$

Constraint (18) ensures that the traffic transmitted through a virtual link does not exceed its maximum capacity.

Flow conservation constraint in the optical layer:

$$\sum_{n \in Nm_m} W_{m,n}^{i,j} - \sum_{n \in Nm_m} W_{m,n}^{i,j} = \begin{cases} C_{i,j} & m = i \\ -C_{i,j} & m = j \\ 0 & otherwise \end{cases}$$

$$\forall \, i, j, m \in N : i \neq j \tag{20}$$

Constraint (20) represents the flow conservation for the optical layer. It ensures that the total number of incoming wavelengths in a virtual link is equal to the total number of outgoing wavelengths in all nodes excluding the source and destination nodes of the virtual link.

Physical link capacity:

$$\sum_{i \in N} \sum_{j \in N : i \neq j} W_{m,n}^{i,j} \leq W \, F_{m,n}$$

$$\forall \, m, n \in N \tag{21}$$

Constraint (22) represents the physical link capacity limit. It ensures that the number of wavelengths in virtual links traversing a physical link does not exceed the maximum capacity of fibres in the physical link.

Total number of aggregation ports in a core node:

$$APC_s = \frac{1}{B} \sum_{d \in N} L_{s,d}$$

$$\forall \, s \in N \tag{22}$$

Constraint (22) calculates the total number of router ports in each core node that aggregate the traffic from/to the clouds.

The mathematical model given above maximizes the total profit of an ISP. To calculate the core network power consumption achieved from the profit-driven model, following parameters and variables are introduced;

Parameters:

$S$    Maximum span distance between two erbium doped fibre amplifiers (EDFAs).

$D_{m,n}$    Distance in kilometres between node pair $(m, n)$.

$A_{m,n}$    Number of EDFAs between node pair $(m, n)$. $A_{m,n} = \left\lfloor \frac{D_{m,n}}{S} - 1 \right\rfloor$ where $S$ is the reach of the EDFA.

$G_{m,n}$    Number of regenerators between node pair $(m, n)$. Typically $G_{m,n} = \left\lfloor \frac{D_{m,n}}{R} - 1 \right\rfloor$, where $R$ is the reach of the regenerator.

$Prp$    Router port power consumption.

$Pt$    Transponder power consumption.
$Pe$    EDFA power consumption.
$Po_s$    Optical switch power consumption in node $s$.
$Prg$    Regenerator power consumption.
$n$    Core network power usage effectiveness.

Under the non-bypass approach [44], the IP over WDM network power consumption is composed of:

The power consumption of routers ports:

$$n\left(\sum_{s\in N} Prp\, APC_s + \sum_{m\in N}\sum_{n\in Nm_m:n\neq m} Prp\, W_{m,n}\right) \quad (22)$$

The power consumption of transponders:

$$n\left(\sum_{m\in N}\sum_{n\in Nm_m:n\neq m} Pt\, W_{m,n}\right) \quad (23)$$

The power consumption of EDFAs:

$$n\left(\sum_{m\in N}\sum_{n\in Nm_m:n\neq m} Pe\, F_{m,n}\, A_{m,n}\right) \quad (24)$$

The power consumption of optical switches:

$$n\left(\sum_{s\in N} Po_s\right) \quad (25)$$

The power consumption of regenerators:

$$n\left(\sum_{m\in N}\sum_{n\in Nm_m:n\neq m} Prg\, RG_{m,n}\, W_{m,n}\right) \quad (26)$$

The total traffic carried over the core physical links is given as:

$$\sum_{m\in N}\sum_{n\in Nm_m:n\neq m} W_{m,n}\, B \quad (27)$$

## IV. PROFIT-DRIVEN MODEL RESULTS:

In this section, we evaluate the increase in ISP profit and the reduction in network traffic and subsequently power consumption resulting from the optimized pricing scheme under the repeal of net neutrality. We define the three services classes as follows;

- Class A; for UHD video service; 18 Mbps download rate.
- Class B; for HD video service; 7.2 Mbps download rate.
- Class C; for SD video service; 2 Mbps download rate.

We investigate CP's end users' choices of service classes based on different PED. We show how users behavior under the different PED; 0.2, 0.4, 0.6, 0.8, 1 or 2 affects the equilibrium price of each class the ISP charges the CP for delivering its content.

As discussed above, we assume that the CP will transfer the price increase to their customers at the same rate (if the CP absorbs some of the increase in prices, then this may represent a different PED). As a benchmark, we consider users to be distributed among classes according to the Cisco forecast report [45], where UHD, HD, and SD users distribution are 19%, 56% and 25% respectively. We consider 1.8 million users active simultaneously in the network. This figure is obtained as follows: The number of users is 44 million users in Netflix in the US and the average user spent around 1 hour per day watching movies in 2015 [46]. Therefore, the average number of users during one hour of the day is 1.8 million users, which is an average number that does not consider the popularity of different viewing times in the day. The concentration of users at any node in AT&T network is based on the population of the state where the node is located (see Fig. 2).

We consider the BT network connectivity selling price as the net neutrality price of the three classes where 10 Gbps connectivity is priced at £12,600 ($15,750) per year [47], i.e. $131 per 1 Gbps link per month. The actual cost of provisioning ISP core network infrastructure is sensitive information and not usually shared by ISPs. However, we estimate the cost of provisioning 1 Gbps of network as $118 considering 10% as the ISP profit margin (the average profit margin for AT&T [2] and Comcast [48] were approximately 9% and 12%, respectively between 2013-2018). We divided the cost among the three network layers; core, metro and access network based on their power consumption percentages: 24%, 6% and 70%, respectively [49] which corresponds to $28, $7, $83, respectively. The cost of $28 per Gbps in the core network is associated with a single hop. For the AT&T architecture the average hop count between clouds and other nodes is 1.

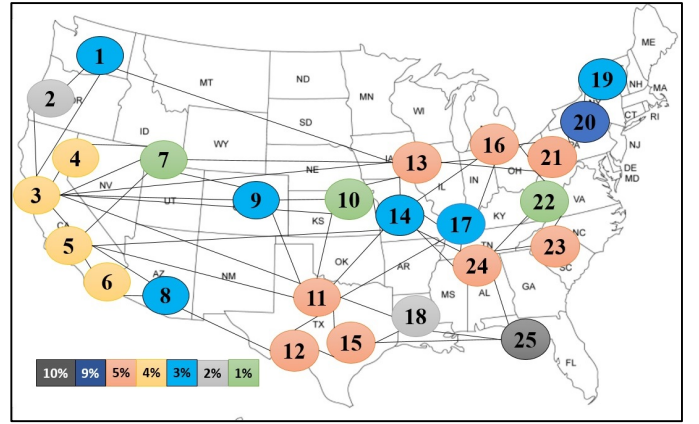

Figure 2: AT&T core network with percentage of population in each node.

As shown in Fig. 2, we choose AT&T core network (a primary core network topology in the US) as a core network topology example. This core network consists of 25 nodes and 54 bidirectional links. AT&T hosts datacenters in nodes 1, 3, 5, 6, 8, 11, 13, 17, 19, 20, 22, and 25 [50]. These nodes are used to host datacenters to serve distributed CPs users. The input parameters used are given in Table I.

TABLE I: INPUT PARAMETERS OF PROFIT-DRIVEN MODEL

| | |
|---|---|
| Router port power consumption ($Prp$) | 638W [51] |
| Transponder power consumption ($Pt$) | 129W [52] |
| Regenerator power consumption ($Prg$) | 114W, reach 2000 km [53] |
| EDFA power consumption ($Pe$) | 11W [54] |

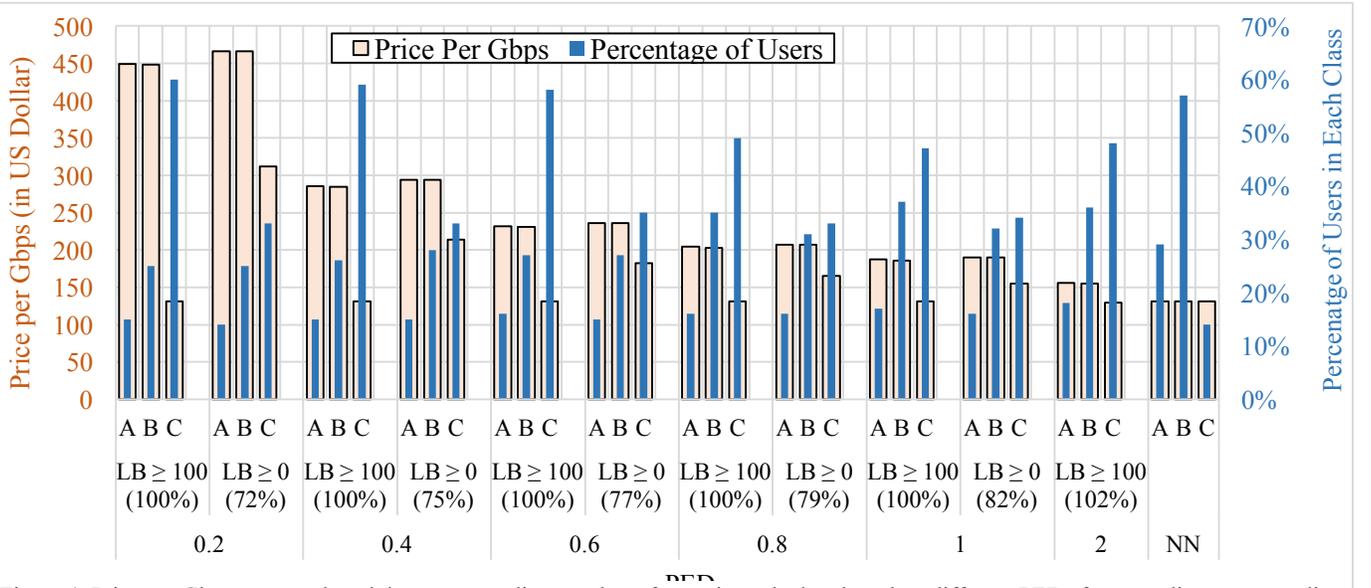

Figure 4: Price per Gbps per month and the corresponding number of users in each class based on different PED after repealing net neutrality (cloud-based delivery).

| | |
|---|---|
| Optical switch power consumption ($Po$) | 85W [55] |
| Number of wavelengths in a fiber (W) | 32 [56] |
| Bit rate of each wavelength (B) | 40 Gbps [56] |
| Span distance between two EDFAs ($S$) | 80 km [54] |
| Network power usage effectiveness ($n$) | 1.5 [57] |
| Total users ($u$) | 1.8 million users [46] |
| The cost of provisioning 1 Gbps of core network bandwidth per month ($\mathbb{C}$) | $28 |
| The cost of provisioning 1 Gbps of metro and access network bandwidth per month ($\mathbb{Э}$). | $90 |
| The net neutrality selling price of downloading 1 Gbps of network bandwidth per month (Ps) | $131 [47] |
| Set of classes ($\alpha$) | 3 classes; A, B and C |
| Number of users of class $i$ located in node $d$ under net neutrality scenario ($N_{d,i}$) | 19% of total users for class A, 56% for class B, and 25% for class C [45]. Number of users in each node is based on the population of the state where the node is located (see Fig. 2). |
| Download rate of class $i$ ($d_i$) | 18 Mbps for class A, 7.2 Mbps for class B, and 2 Mbps for class C [45] |
| Price elasticity of demand ($E_i$) | 0.2, 0.4, 0.6, 0.8, 1 or 2 |
| Minimum percentage of users served by CP to be maintained by the pricing scheme ($LB$). | 0 or 100 |

In the following subsections, we evaluate two scenarios; equal PED for all classes and different PED for different classes.

Under each scenario we study three scenarios of delivering CPs contents to users; a cloud-based delivery and a cloud-fog based deliver and fog-based delivery.

*1) Equal PED among classes:*

In the following, we study three scenarios of delivering CPs contents to users; a cloud-based delivery and a cloud-fog based delivery and fog-based delivery.

**Cloud based delivery:** Figs. 3 to 5 show the profit-driven model results for AT&T core network where content is delivered from the 12 datacenters in the AT&T topology [50]. The number of users and the corresponding price of each class under different PED are illustrated in Fig. 3. The primary y-axis shows price per Gbps per month of each class in US dollar. These prices represent the equilibrium point of users' willingness to follow the price increase which results in maximum profit for the ISP. The secondary y-axis corresponds to the percentage of users subscribed to each class. The x-axis shows different PED scenarios from 2 to 0.2. The former represents the highest sensitivity to the price change considered, whereas, the latter represents the contrary. PED values are shown along with the case of net neutrality where the price of different classes is fixed at 113$ and the percentage of users in each class follows Cisco forecast report [45] as discussed above. For each PED value we consider two cases; a case where the optimized pricing scheme should maintain 100% of the users that existed under net neutrality ($LB \geq 100$) and another case where the pricing scheme can result in users leaving the service ($LB \geq 0$).

Fig. 4 is a plot of the monthly profit of ISP considering different PED values as well as net neutrality scenario. Total traffic of core network and the power consumption due to this traffic under different PED scenarios and the net neutrality scenario are plotted in Fig. 5. In case of content with $PED = 2$, under $LB \geq 100$ or $LB \geq 0$, Fig. 3 shows that repealing net neutrality has increased class C users to 48% of the total number of users compared to 14% only under the net neutrality pricing scheme. This increase is a result of some users of class B downgrading to class C as the class B price increased slightly by 18% (the number of users in class B reduced to 36%) and due to new users joining the service (the total number of users increased to 102%) attracted by the 1% decrease in class C price. The users of class A are reduced to 18% of the total number of users as a result of the slight increase in price by 19%. This pricing scheme and distribution of users have resulted in an increase in the total profit by 54% compared to the net neutrality scenario as seen in Fig. 4. For a less sensitive content with $PED = 0.2$ under $LB \geq 0$, the equilibrium pricing scheme resulted in 28% of the users leaving the service as the increase in the classes price resulted in an increase in the profit by a factor of 8.3 compared to the net neutrality scenario. Maintaining all the users of the service ($LB \geq 100$) has slightly reduced the profit by 10%.

In addition to growing ISP profit, we also observe in Fig. 5 a decline in the core network traffic by up to 55% under $PED = 0.2$, $LB \geq 0$ and a consequent reduction in power consumption by 49%. This reduction in core network traffic and power consumption occurred for two reasons; 1) some cloud service users leave classes A and B to subscribe to class C as the charges per Gb/s of the classes A and B increase. 2) the total cloud service subscribers diminished due to the increase in class C price (in case of $LB \geq 0$).

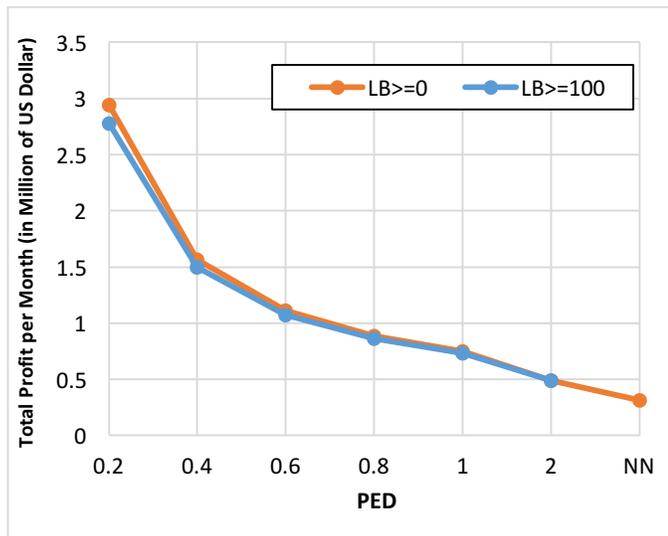

Figure 5: Total profit per month of profit-driven model under different PED scenarios for cloud-based delivery.

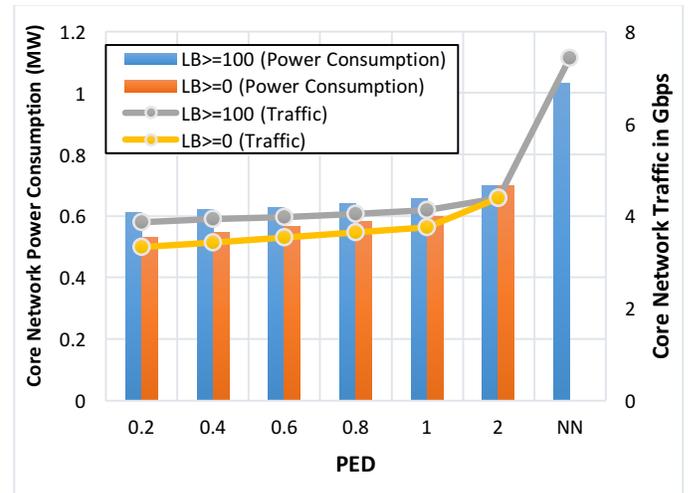

Figure 6: Total core network traffic and power consumption of profit-driven model under different PED scenarios for cloud-based delivery.

**Cloud-Fog based delivery:** Next, we introduce 10 fog nodes in addition to the 12 datacenter locations. These fog nodes are assumed to be built in the proximity of nodes with the highest population in the AT&T core network, so no core network cost (₡) is incurred by serving the demands of these nodes. Fig. 6 shows that the prices per Gbps per month under different PED that are less than the previous case (cloud-based delivery) as we reduced the cost of the core network by introducing the fog nodes. Under $PED = 2$, the prices compared to the net neutrality case in class A and B increased by 12% and 11%, respectively, while the price of class C dropped by 1% as opposed to 19%, 18% and 1% with cloud-based delivery. The reduced prices attracted more users resulting in increase in the profit by 18% compared to the net neutrality case as seen in Fig. 7 as opposed to a 54% increase in profit with cloud-based delivery. Fig. 8 shows a reduction in core network traffic (40%) and power consumption (35%) by repealing net neutrality in the cloud-fog architecture.

**Fog based delivery:** Here, we consider a scenario in which all users access CP contents from a local fog node. Although deploying a fog node locally, to serve CP customers, increases the capital expenditure (CAPEX) and operating expenses (OPEX) of provisioning multiple locations (i.e. 25 fog nodes in AT&T network), it reduces the communication network transit cost burden to the minimum. However, fog nodes are not always an option due to the finite capacity of processing and storage. The results show that the prices are further reduced under fog-based delivery (Fig. 9) as no core network cost (₡) is incurred by serving demands. For instance, under $PED = 2$, the prices compared to the net neutrality case in class A and B increased by 9% while the price of class C is decreased by 11% resulting in increase in the profit by 6% compared to the net neutrality scenario as seen in Fig. 10.

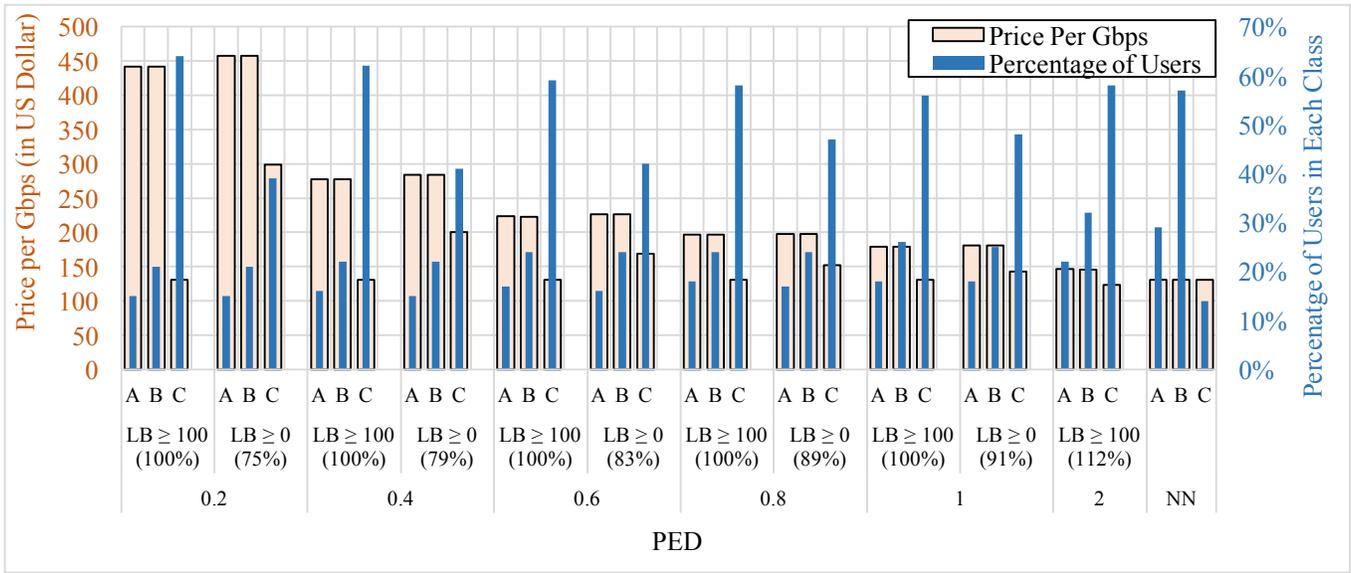

Figure 7: Price per Gbps per month and the corresponding number of users in each class based on different PED after repealing net neutrality (cloud-fog based delivery).

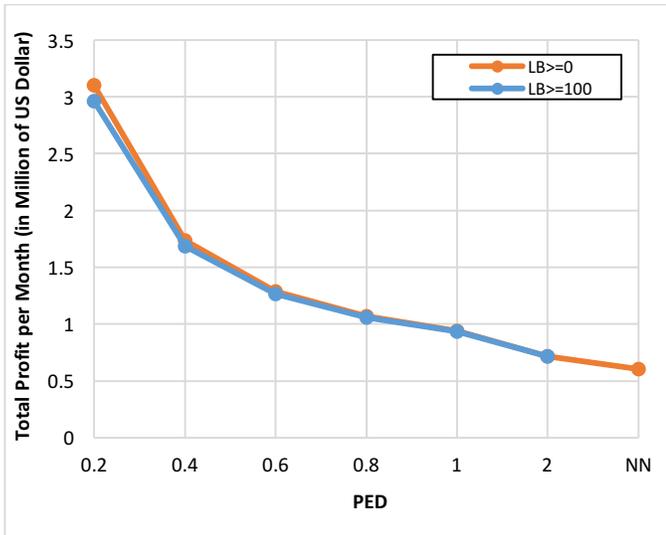

Figure 8: Total profit per month of profit-driven model under different PED (cloud-fog based delivery).

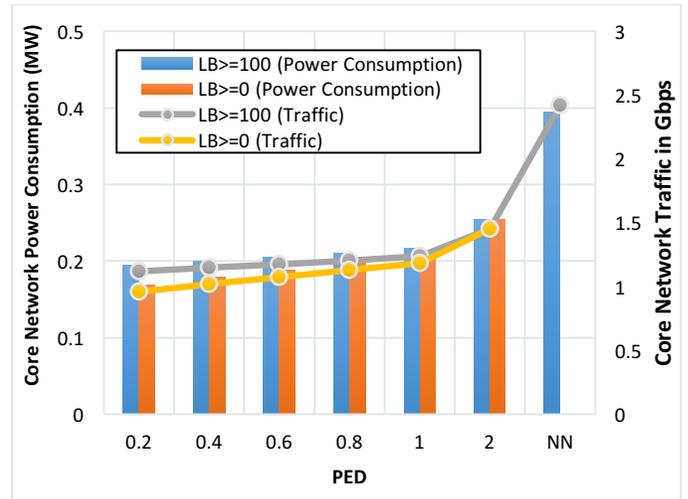

Figure 9: Total core network power consumption and traffic of profit-driven model under different PED (cloud-fog based delivery).

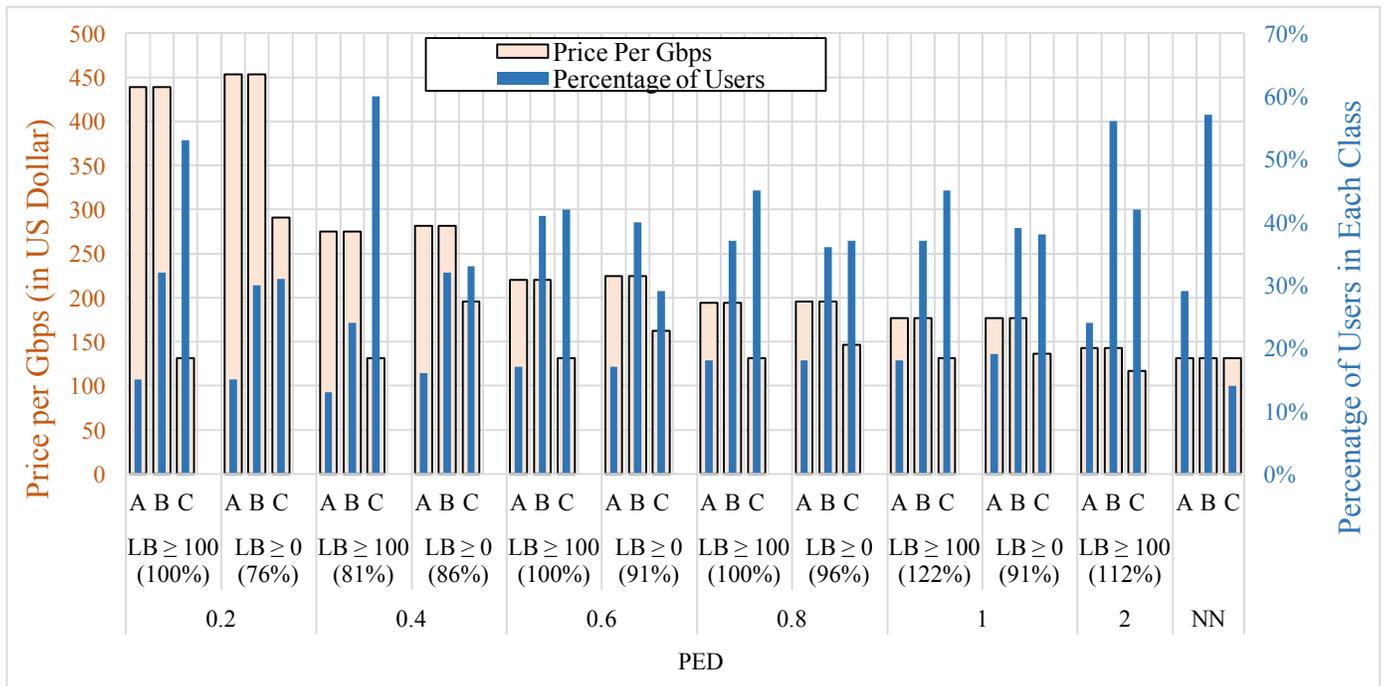

Figure 10: Price per Gbps per month and the corresponding number of users in each class based on different PED after repealing net neutrality (fog-based delivery).

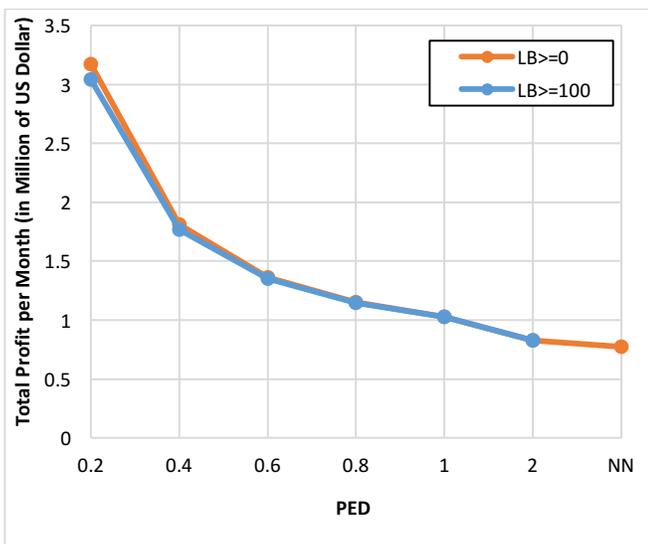

Figure 11: Total profit per month of profit-driven model under different PED (fog-based delivery).

*2) Different PED among classes:*

In this section, we consider a scenario where elasticity of demand varies among the different classes of service. We consider class C to be less sensitive to price change than class B. Also, we considered class B to be less sensitive than class A. The elasticity of demand for classes A, B and C are considered to be 2, 0.8 and 0.2, respectively. Fig. 11 shows the price per Gbps for classes A and B is the same under different scenarios and delivery schemes as a result of the high PED of class A. Class C is priced at the same level of classes A and B for $LB \geq 0$ as the low PED of class C limits the number of users leaving the services as a result of increase in the price. Fig. 12 shows an increase in profit by up to 88%, 29% and 16% under cloud-based delivery, cloud-fog based delivery and fog-based delivery, respectively, compared to the net neutrality scenario. Fig 13 shows a decrease in core network traffic by up to 43% and 30% under cloud-based delivery and cloud-fog based delivery, respectively, compared to the net neutrality scenario. Also, the total reduction in the core network power consumption (as shown in Fig 14) is up to 40% and 32% under cloud-based delivery and cloud-fog based delivery respectively.

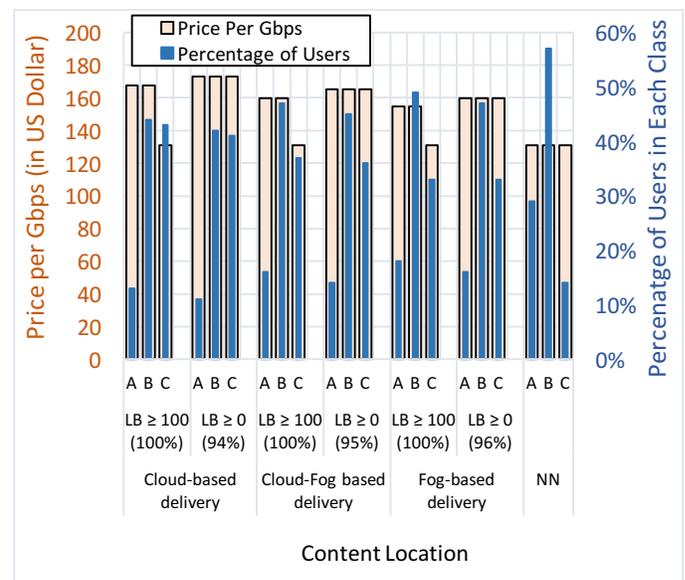

Figure 12: Price per Gbps per month and the corresponding number of users in each class of profit-driven model for different CP delivery scenarios where PED values of different classes A, B and C are 2, 0.8 and 0.2, respectively.

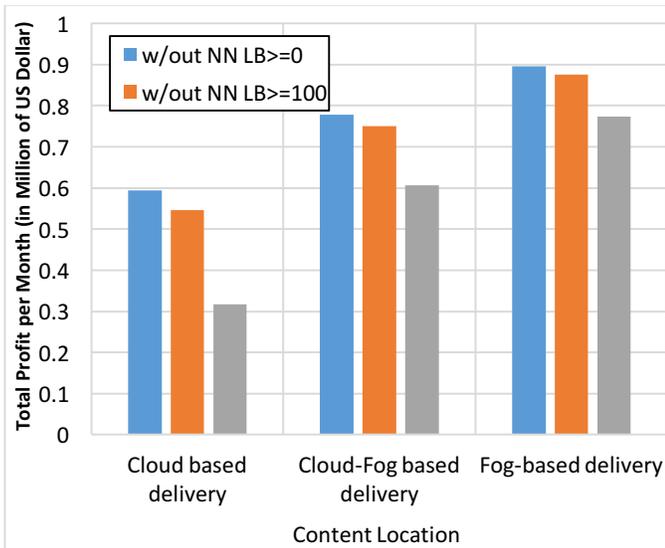

Figure 13: Total profit per month of profit-driven model for different CP delivery scenarios where PED values of different classes A, B and C are 2, 0.8 and 0.2, respectively.

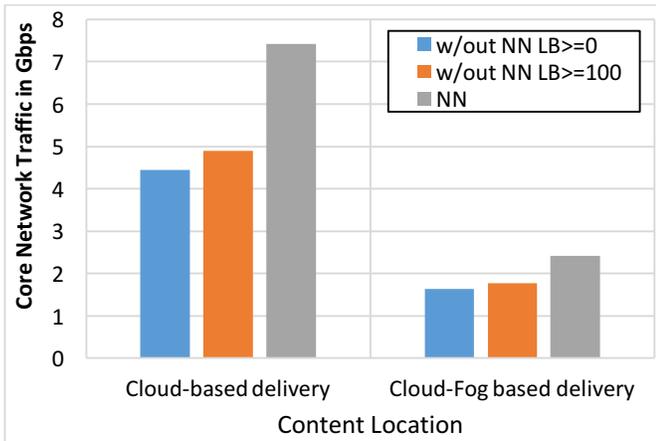

Figure 13: Total traffic resulting from profit-driven model for different CP delivery scenarios where PED values of different classes A, B and C are 2, 0.8 and 0.2, respectively.

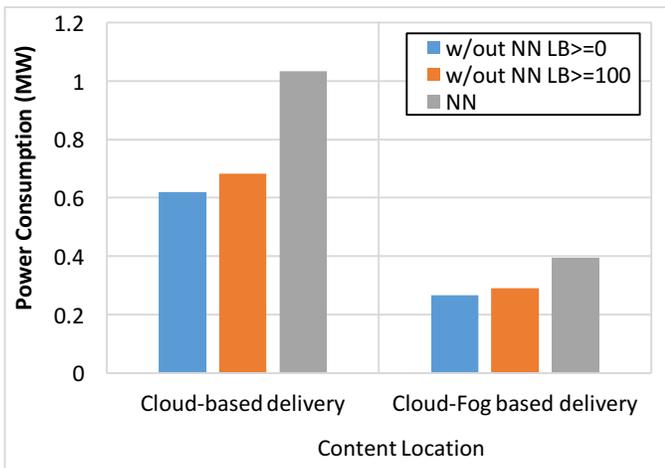

Figure 14: total core network power consumption resulting from profit-driven model for different CP delivery scenarios where PED values of different classes A, B and C are 2, 0.8 and 0.2, respectively.

## V. CONCLUSIONS

In this paper, we developed a MILP model to optimize the pricing scheme used by ISPs to charge CPs for delivering their video content under the repeal of net neutrality where ISPs can treat data intensive traffic less favorably. A techno-economic Mixed Integer Linear Programming (MILP) model is developed to maximize the ISP profit by optimizing the ISP pricing scheme to charge different classes of service differently subject to PED. We considered three classes of service that represent different data rate requirements of video content. The analysis addressed three CP delivery scenarios; cloud-based delivery, cloud-fog based delivery and fog-based delivery. The results show that the discriminatory pricing scheme can increase the ISPs profit by a factor of 8. The results also show that by influencing the way end-users consume data-intensive content, the core network traffic and consequently power consumption are reduced by up to 49% and 55%, respectively, compared to the net neutrality scenario.

**Acknowledgements**

The authors would like to acknowledge funding from the Engineering and Physical Sciences Research Council (EPSRC), INTERNET (EP/H040536/1), STAR (EP/K016873/1) and TOWS (EP/S016570/1) projects. The first author would like to acknowledge the Government of Saudi Arabia and Taibah University for funding his PhD scholarship. All data are provided in full in the results section of this paper.

## Biographies

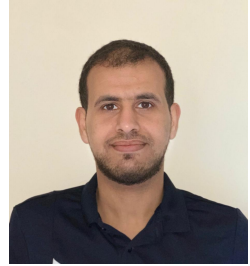

**Hatem A. Alharbi** received the B.Sc. degree in Computer Engineering (Hons.) from Umm Alqura University, Makkah, Saudi Arabia in 2012, the M.Sc. degree in Digital communication networks (with distinction) from University of Leeds, United Kingdom, in 2015. He is currently pursuing the Ph.D. degree with the School of Electronic and Electrical Engineering, University of Leeds, UK. He is currently a Lecturer in Computer Engineering department in the School of Computer Science and Engineering, University of Taibah, Saudi Arabia.

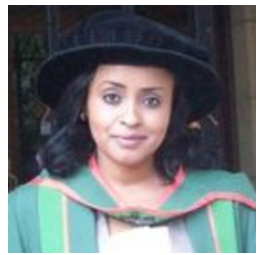

**Taisir EL-Gorashi** received the B.S. degree (first-class Hons.) in electrical and electronic engineering from the University of Khartoum, Khartoum, Sudan, in 2004, the M.Sc. degree (with distinction) in photonic and communication systems from the University of Wales, Swansea, UK, in 2005, and the PhD degree in optical networking from the University of Leeds, Leeds, UK, in 2010. She is currently a Lecturer in optical networks in the School of Electrical and Electronic Engineering, University of Leeds. Previously, she held a Postdoctoral Research post at the University of Leeds (2010– 2014), where she focused on the energy efficiency of optical networks investigating the use of renewable energy in core networks, green IP over WDM networks with datacenters, energy efficient physical topology design, energy efficiency of content distribution networks, distributed cloud computing, network virtualization and Big Data. In 2012, she was a BT Research Fellow, where she developed energy efficient hybrid wireless-optical broadband access networks and explored the dynamics of TV viewing behavior and program popularity. The energy efficiency techniques developed during her postdoctoral research contributed 3 out of the 8 carefully chosen core network energy efficiency improvement measures recommended by the GreenTouch consortium for every operator network worldwide. Her work led to several invited talks at GreenTouch, Bell Labs, Optical Network Design and Modelling conference, Optical Fiber Communications conference, International Conference on Computer Communications, EU Future Internet Assembly, IEEE Sustainable ICT Summit and IEEE 5G World Forum and collaboration with Nokia and Huawei.

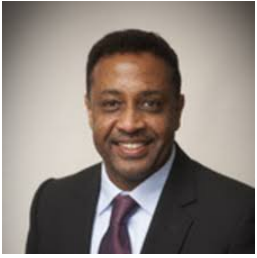**Jaafar M. H. Elmirghani** is the Director of the Institute of Communication and Power Networks within the School of Electronic and Electrical Engineering, University of Leeds, UK. He joined Leeds in 2007 and prior to that (2000–2007) as chair in optical communications at the University of Wales Swansea he founded, developed and directed the Institute of Advanced Telecommunications and the Technium Digital (TD), a technology incubator/spin-off hub. He has provided outstanding leadership in a number of large research projects at the IAT and TD. He received the Ph.D. in the synchronization of optical systems and optical receiver design from the University of Huddersfield UK in 1994 and the DSc in Communication Systems and Networks from University of Leeds, UK, in 2014. He has co-authored Photonic switching Technology: Systems and Networks, (Wiley) and has published over 500 papers. He has research interests in optical systems and networks. Prof. Elmirghani is Fellow of the IET, Fellow of the Institute of Physics and Senior Member of IEEE. He was Chairman of IEEE Comsoc Transmission Access and Optical Systems technical committee and was Chairman of IEEE Comsoc Signal Processing and Communications Electronics technical committee, and an editor of IEEE Communications Magazine. He was founding Chair of the Advanced Signal Processing for Communication Symposium which started at IEEE GLOBECOM'99 and has continued since at every ICC and GLOBECOM. Prof. Elmirghani was also founding Chair of the first IEEE ICC/GLOBECOM optical symposium at GLOBECOM'00, the Future Photonic Network Technologies, Architectures and Protocols Symposium. He chaired this Symposium, which continues to date under different names. He was the founding chair of the first Green Track at ICC/GLOBECOM at GLOBECOM 2011, and is Chair of the IEEE Sustainable ICT Initiative within the IEEE Technical Activities Board (TAB) Future Directions Committee (FDC) and within the IEEE Communications Society, a pan IEEE Societies Initiative responsible for Green and Sustainable ICT activities across IEEE, 2012-present. He is and has been on the technical program committee of 38 IEEE ICC/GLOBECOM conferences between 1995 and 2019 including 18 times as Symposium Chair. He received the IEEE Communications Society Hal Sobol award, the IEEE Comsoc Chapter Achievement award for excellence in chapter activities (both in 2005), the University of Wales Swansea Outstanding Research Achievement Award, 2006, the IEEE Communications Society Signal Processing and Communication Electronics outstanding service award, 2009, a best paper award at IEEE ICC'2013, the IEEE Comsoc Transmission Access and Optical Systems outstanding Service award 2015 in recognition of "Leadership and Contributions to the Area of Green Communications", received the GreenTouch 1000x award in 2015 for "pioneering research contributions to the field of energy efficiency in telecommunications", the 2016 IET Optoelectronics Premium Award and shared with 6 GreenTouch innovators the 2016 Edison Award in the "Collective Disruption" Category for their work on the GreenMeter, an international competition, clear evidence of his seminal contributions to Green Communications which have a lasting impact on the environment (green) and society. He is currently an editor of: IET Optoelectronics, Journal of Optical Communications, IEEE Communications Surveys and Tutorials and IEEE Journal on Selected Areas in Communications series on Green Communications and Networking. He was Co-Chair of the GreenTouch Wired, Core and Access Networks Working Group, an adviser to the Commonwealth Scholarship Commission, member of the Royal Society International Joint Projects Panel and member of the Engineering and Physical Sciences Research Council (EPSRC) College. He was Principal Investigator (PI) of the £6m EPSRC INTelligent Energy awaRe NETworks (INTERNET) Programme Grant, 2010-2016 and is currently PI of the £6.6m EPSRC Terabit Bidirectional Multi-user Optical Wireless System (TOWS) for 6G LiFi Programme Grant, 2019-2024. He has been awarded in excess of £30 million in grants to date from EPSRC, the EU and industry and has held prestigious fellowships funded by the Royal Society and by BT. He was an IEEE Comsoc Distinguished Lecturer 2013-2016.